\documentclass[aps,prd,twocolumn,groupedaddress,showpacs]{revtex4-1}
\usepackage{amsmath,amstext,amsbsy,amssymb}
\usepackage{bm}
\usepackage{color}

\newcommand{\diag}{{\rm diag}}

\newcommand{\cme}{{\scriptscriptstyle{ CME}}}

\newcommand{\cve}{{\scriptscriptstyle{ CVE}}}
\newcommand{\ssA}{{\scriptscriptstyle{ A}}}
\newcommand{\ssV}{{\scriptscriptstyle{ V}}}

\newcommand{\ssS}{{\scriptscriptstyle{ S}}}

\newcommand{\ssE}{{\scriptscriptstyle{ E}}}
\newcommand{\ssB}{{\scriptscriptstyle{ B}}}
\newcommand{\ssL}{{\scriptscriptstyle{ L}}}

\newcommand{\ssC}{{\scriptscriptstyle{ C}}}
\newcommand{\ssP}{{\scriptscriptstyle{ P}}}

\newcommand{\sschi}{{\scriptscriptstyle{ \chi}}}

\newcommand{\ssfd}{{\scriptscriptstyle{FD}}}
\newcommand{\ssBb}{{\scriptscriptstyle{ B5}}}
\newcommand{\ssbes}{{\scriptscriptstyle{ 5}}}

\newcommand{\ssmn}{{\scriptscriptstyle{\mu \nu }}}

\newcommand{\ssbfn}{{\scriptscriptstyle{(n)}}}
\newcommand{\ssbfu}{{\scriptscriptstyle{(u)}}}

\long\def\symbolfootnote[#1]#2{\begingroup%
\def\thefootnote{\fnsymbol{footnote}}\footnote[#1]{#2}\endgroup}

\begin{document}
	
\title{Quantum Kinetic Equation in the Rotating Frame and Chiral Kinetic Theory}

\author{\"{O}mer F. Dayi}
\author{Eda Kilin\c{c}arslan}
\affiliation{%
	Physics Engineering Department, Faculty of Science and
	Letters, Istanbul Technical University,
	TR-34469, Maslak--Istanbul, Turkey}

\begin{abstract}
 A modified quantum kinetic equation which  takes account of the  noninertial features  of   rotating frame is proposed. The vector and axial-vector field components of the Wigner function  for chiral fluids are worked out
  in a semiclassical scheme.  It is demonstrated that the chiral currents and energy-momentum tensor computed by means of them  are consistent with  the hydrodynamical  results.   A new semiclassical  covariant  chiral transport equation is established by inspecting  the equations satisfied by the chiral vector fields.   It uniquely provides a new  three-dimensional semiclassical chiral kinetic  theory  possessing a  Coriolis  force term.  The particle number and current densities deduced from this transport equation  satisfy  the anomalous continuity equation and generate the magnetic and vortical effects correctly.	
\end{abstract}

\maketitle

\section{Introduction}   
Chiral fermions appear in different sorts of physical systems like heavy-ion collisions  \cite{kmw,fkw},  electroweak plasma in early Universe  \cite{js,tvv,bp} and Weyl semimetals in condensed matter physics  \cite{ws1,ws2,ws3}.  In these systems collective  dynamical properties are affected by chiral anomalies which yield the chiral magnetic effect \cite{kmw,fkw,kz}, the  chiral separation effect \cite{mz,jkr},  the chiral vortical effect \cite{ss} and the local (spin) polarization effect \cite{lw,bpr,glpww}.  The latter two are due to the rotation of reference frame. Chiral particles have  been studied within the relativistic hydrodynamical approach in  \cite{ss,sadofyev, kharzhdy,neiman}.
Dynamics of relativistic fluids can also be investigated by means of the quantum kinetic equation (QKE)
\begin{equation}
\gamma_\mu \left(p^\mu + \frac{i\hbar}{2} \nabla^\mu\right) W(x,p) = 0,
\label{qkeO}
\end{equation}
 derived from the Dirac equation coupled to electromagnetic fields \cite{qBe,vge},  in the phase space described  by the position and momentum four-vectors $x_\mu,p_\mu.$ 
We consider a  constant  electromagnetic field strength $F_{\mu \nu },$ so that
$\nabla^\mu =\partial^\mu - Q F^{\mu\nu}   \partial_\nu^p , $ where $ \partial^\mu  \equiv \partial / \partial x_\mu,\   \partial_\mu^p  \equiv \partial / \partial p^\mu.$ $W(x,p)$  is the 
Wigner function for spin-$1/2$ fermions which  can be  decomposed in terms of sixteen independent generators of the Clifford algebra
whose coefficients are  scalar, pseudoscalar, vector,  axial-vector  and tensor fields. 
When fermions are massless  the vector and axial-vector fields satisfy a set of coupled equations.
 In   \cite{glpww} chiral vector   fields   were established semiclassically by considering the chiral fluid in the vicinity of local equilibrium. Thus, $x_\mu$ dependence of the fields is due to the  fluid four-velocity $u_\mu (x).$ Then the derivatives of fields generate terms  depending on the  fluid vorticity $\omega^\mu= (1/2) \epsilon^{\mu\nu\alpha \rho} u_\nu \partial_\alpha u_{\rho} .$   

QKE  may also be employed to designate relativistic semiclassical chiral kinetic theory (CKT).  Kinetic theory, needless to say, is essential to study the dynamics of chiral quark plasma. We consider only collisionless kinetic theories.
There are mainly two different Lorentz covariant approaches which employ QKE: In  \cite{cpww}  CKT was established by acting $\nabla_\mu$ on the vector and axial-vector fields 
derived in  \cite{glpww}  where  the momentum dependence of   distribution function  is only due to the  component  parallel to $u_\mu,$ namely $ u\cdot p.$   The vorticity dependent terms acquired in \cite{cpww} do not lead the desired currents and energy-momentum  tensor.  To surmount this difficulty  one should add some terms to the time evolutions of phase space variables \cite{gpw}. Moreover, in this method three-dimensional (3D) semiclassical CKT  is not uniquely determined.  
The other relativistic approach \cite{hpy1,hpy2} is based on the chiral vector fields computed perturbatively within the quantum field theory. These  can also be deduced from (\ref{qkeO}) by studying its solution in a frame moving with the velocity $n_\mu$ \cite{hsjlz}.  A covariant CKT is provided by acting  $\nabla_{\mu}$ on them. Unfortunately, 3D limit of this relativistic equation for nonvanishing  vorticity  has not been discussed.  Different from these, there is  a  worldline   formalism approach to CKT  \cite{mv}.

In \cite{sy} it was claimed that the vorticity appearing in the hydrodynamical approach matches the angular velocity of the fluids in the comoving frame.  In fact, they  acquired the chiral vortical effect in 3D from the chiral magnetic effect by substituting magnetic field  with angular velocity times energy suggested by the similarity of  the Coriolis and  Lorentz forces. This statement inspired the   explicitly   spatial coordinate dependent  3D   semiclassical kinetic theory formulation of fermions  presented in \cite{dky}.
In this method the Coriolis force appears in contrary to  the  formulations of \cite{cpww, hpy2, hsjlz} based on QKE. 

Here we  follow a different route. The QKE derived from the Dirac equation including  electromagnetic interactions is frame independent.  In the absence of background fields it cannot yield a transport equation (TE) possessing  terms  which can be interpreted as force,  so that non-inertial forces like the Coriolis force do not appear.  We propose a modification of (\ref{qkeO})  by making use of enthalpy current, such that  the modified QKE suits well with the noninertial  properties of the frame.  Although, the modified QKE is valid for fermions either  massive or massless, we  only deal  with chiral fluids. We first present the set of equations for chiral vector fields and  discuss their  semiclassical  solutions by a general distribution function. 
To ensure  the consistency of the modified QKE we present its solution for chiral vector fields in the comoving frame of fluid by choosing a specific  distribution function and  show that they lead to  the correct vector, axial-vector currents and energy-momentum tensor.   
Then by employing the chiral vector fields written in terms of general distribution function, 
a covariant chiral transport equation (CTE) is established in the comoving frame of the fluid.  
Procedure of deriving the 3D limit of a four-dimensional TE by integrating it over $p_0$ is  clarified and applied to the covariant CTE.  We showed that  a unique  3D CKT follows. 
The phase space measure, velocity and force imposed by this new CKT generate the desired continuity equations, magnetic effects,  vortical effects as well as  the Coriolis  force.  Our results  demonstrate  that the  modified QKE is  reliable and the proposed modification is essential to   satisfactorily address  the noninertial features of the rotating  frame.

\section{Modified Quantum Kinetic Equation}
  
The fluid four-velocity $u_\mu= d x_\mu /d\tau,$  is
defined by  the proper time $\tau,$ so that it satisfies $u_\mu u^\mu=1.$
The metric tensor is $g_{\mu\nu}= \diag (1,-1,-1,-1).$  
One can split a four-vector  $y_\mu$ in directions parallel and perpendicular to $u_\mu$ as
 $
 y_\mu =y_0u_\mu +\bar{y}_\mu,
 $
 where $y_0\equiv y\cdot u  $ and $\bar{y}_\mu\equiv (g_{\mu\nu}-u_\mu u_\nu )y^\nu .$ The  (non)inertial properties of relativistic vorticity are provided by the  circulation tensor \cite{ReZo}
 $$
 W_{\mu \nu}= \partial_\mu W_\nu - \partial_\nu W_\mu ,
 $$
 where $W_\mu =h u_\mu$ is the enthalpy current given in terms of the internal energy (enthalpy) $h=p\cdot u.$   
For  fluids which are not subject to acceleration  $a_\mu= u_\nu \partial^\nu u_\mu =0,$  the circulation tensor can be expressed as 
 $$
 W_{\mu \nu}= 2h \Omega_{\mu \nu}+(\partial_\mu h)u_\nu - (\partial_\nu h)u_\mu,
 $$
 where 
 $\Omega_{\mu \nu}=\frac{1}{2} \left(\partial_\mu u_\nu - \partial_\nu u_\mu \right)$ is  the (kinematic) vorticity  tensor.
  Observe that  the  force acting on a fluid element imposed by  (\ref{qkeO})  is characterized by the coefficient of the derivative with respect to momentum four-vector and there is an apparent resemblance between  $F_{\mu \nu}$ and $W_{\mu \nu}.$  Hence, to take into account the noninertial forces one is tempted  to modify (\ref{qkeO}) by adding a term proportional to $W^{\mu \nu}\partial^p_\nu$ to $\nabla^\mu.$  However in the rest frame of massive particles where $h$ is constant, the Coriolis force should vanish. 
  Therefore, we add $ \left(W^{\mu \nu}-2h \Omega^{\mu \nu}\right)\partial^p_\nu$  to $\nabla^\mu$ to take into account the noninertial forces.  Generalizing this consideration to a frame  moving with the four-velocity $n_\mu,$ satisfying $n_\mu n^\mu =1,$ we define  
 $$\tilde{\nabla}_{\ssbfn}^\mu\equiv \partial_x ^\mu - \left[ Q F^{\mu\nu}  + (\partial^\mu n^\alpha )  p_\alpha  n^\nu - (\partial^\nu n^\alpha )  p_\alpha  n^\mu \right] \partial_\nu^p .$$
Therefore, for a constant field strength $F^{\mu\nu}$ we propose  
\begin{equation}
\gamma_\mu (p^\mu + \frac{i\hbar}{2} \tilde{\nabla}_{\ssbfn}^\mu) W(x,p) = 0,
\label{qke}
\end{equation}
as the  QKE in the rotating frame of reference. 

There are  similarities between the proposed modification and  shifting gauge  fields \cite{ssz,ban} to deal with moving  medium. Our approach resembles  the effective field theory formalism given in  \cite{ssz}  where the gauge field is shifted by the fluid velocity multiplied by chemical potential.  However, treating the enthalpy current as a gauge field does not yield the desired modification. Even if one can   introduce a  gauge field like function to generate the modification,  it should be proportional to  internal energy which   depends on momentum. This  would create problems in Lagrange formulation defined in terms of  spacetime coordinates and derivatives with respect to them.

We are interested in  the chiral vector fields 
\begin{equation}
{\cal J}^\mu_\sschi = \frac{1}{2} ({\cal V}^\mu + \chi {\cal A}^\mu), \nonumber
\end{equation} 
constructed by the vector and axial-vector field  components  ${\cal V}_\mu $  and  ${\cal A}_\mu$  of $W(x,p).$ They correspond to the right-handed $\chi =1,$  and left-handed $\chi =-1,$  fermions.
The modified QKE  (\ref{qke}), yields the following set of  equations
\begin{eqnarray}
p_\mu {\cal J}_\sschi^\mu & = & 0,
\label{1st,0} \\
\tilde{\nabla}_{\ssbfn}^\mu {\cal J}_{\sschi \mu }& = & 0,
\label{2nd,0}\\
\hbar \epsilon_{\mu \nu \alpha \rho} \tilde{\nabla}_{\ssbfn}^\alpha {\cal J}^\rho_\sschi&=& - 2 \chi (p_\mu {\cal J}_{\sschi \nu} -  p_\nu {\cal J}_{\sschi\mu}) ,
\label{third,0}
\end{eqnarray}
which are decoupled  from  the other components.

\section{Semiclassical approximation}
  
To study  solutions of (\ref{1st,0})-(\ref{third,0}) we expand  ${\cal J}^\mu_\sschi$  in $\hbar$ and keep  the zeroth- and  first-order terms \cite{glpww}:
${\cal J}^\mu_\sschi ={\cal J}^{(0)\mu}_\sschi+ \hbar  {\cal J}^{(1)\mu}_\sschi.$
The zeroth-order solution of  (\ref{1st,0}) and (\ref{third,0}) is 
\begin{equation}
{\cal J}^{(0)\mu}_\sschi= p^\mu \delta(p^2) f^{0}_\sschi.
\label{J0}
\end{equation}
$f^{0}_\sschi$ is a general distribution function which  can be decomposed into   the particle  and antiparticle parts,  $s=\pm 1,$  as
$f^{0}_\sschi = \sum_{s=\pm 1} \theta(s  n\cdot p)  f^{0}_{s, \sschi} (x, p).  $ 

At the $\hbar$-order   (\ref{1st,0}) and (\ref{third,0}) lead to
\begin{eqnarray}
p^\mu {\cal J}^{(1)}_{\sschi\mu} &= &0,
\label{1st,1}\\
\epsilon^{\mu \nu \alpha \rho} \tilde{\nabla}_{\ssbfn \alpha } {\cal J}^{(0)}_{\sschi\rho}  &= &- 2  \chi [p^\mu {\cal J}^{(1)\nu}_\sschi-  p^\nu {\cal J}^{(1)\mu}_\sschi ]
\label{third,1}. 
\end{eqnarray}
The general form of ${\cal J}^{(1)}_{\sschi\mu}$  satisfying   (\ref{1st,1}) and (\ref{third,1}) is 
\begin{eqnarray}
 {\cal J}^{(1)\mu}_{\sschi} &=& p^\mu f^{1}_\sschi \delta(p^2) + \frac{1}{2}\chi Q \epsilon^{\mu \nu \alpha \beta}F_{\alpha \beta}  p_\nu f^{0}_{\sschi} \delta^\prime (p^2)  \nonumber\\
 &+&  \chi  \epsilon^{\mu \nu \alpha \rho}  p_\nu (\partial_\alpha n_{\beta} ) p^\beta  n_\rho  f^{0}_{\sschi} \delta^\prime (p^2) + \mathcal {K^\mu}  , 
 \label{generalform}
 \end{eqnarray}
 where $ \delta^\prime (p^2) = - \delta(p^2)/p^2$ and $ f^{1}_\sschi $ is a general function which can be considered as the first-order part of the   distribution function: $f_\chi\equiv  f^{0}_\sschi +\hbar  f^{1}_\sschi . $ Inspired by  the results of \cite{hpy1,hpy2,hsjlz}  we  obtain  $\mathcal {K^\mu}$ satisfying (\ref{1st,1}) and (\ref{third,1}) as
$$ \mathcal {K^\mu}  =  S^{\mu \nu}_\ssbfn (\tilde{\nabla}_{\ssbfn \nu} f^{0}_{\sschi}) \delta(p^2), $$
where
$$
S^{\mu \nu}_\ssbfn =
\frac{\chi}{ 2 n \cdot p}  \epsilon^{\mu \nu \rho \sigma} p_\rho  n_\sigma  
$$ 
corresponds to spin. Distribution function should be chosen consistently to satisfy the remaining equation (\ref{2nd,0}). This will lead to the definition of covariant semiclassical chiral equation.  However, let us first demonstrate that there exists  a solution of  (\ref{1st,0})-(\ref{third,0}) yielding results which are  consistent with the ones existing in the literature. 

 \section{Fermi-Dirac distribution}
   
 We would like to present a specific  solution in the comoving frame of the fluid $n_\mu=u_\mu,$ by choosing the distribution function $ f^{0}_{\sschi}$ as in  \cite{glpww}:
 \begin{equation}
 f^{\scriptscriptstyle{FD}}_{\sschi} = \frac{2}{(2 \pi \hbar)^3}  \sum_{s=\pm 1}    \frac{\theta(s  n\cdot p)}{ e^{s( u \cdot p - \mu_\sschi) / T } +1  } \cdot
 \label{f_fd}
 \end{equation}
The chemical potentials of chiral particles
  $\mu_\sschi ,$  are given in terms of the total and chiral chemical potentials as $\mu_{\scriptscriptstyle{R,L}} = \mu \pm \mu_5.$ Note that one ignores the derivatives of theta functions which yield vanishing contribution when one performs  four-momentum  integrals to calculate physical quantities \cite{glpww}.  
  ${\cal J}^{{\ssfd} (0)  \mu}_\sschi= p^\mu \delta(p^2) f^{\scriptscriptstyle{FD}}_{\sschi} $
satisfies  (\ref{2nd,0}) for constant temperature $T,$ by letting
\begin{eqnarray}
 \partial_\alpha \mu  = - Q E_\alpha ,\  \partial_\alpha \mu_5=0, \ \partial_{\mu}u_\nu =-\partial_{\nu}u_\mu .
\label{emuT}
\end{eqnarray}
$E_\mu = F_{\mu \nu }u^\nu$  is the electric field. The magnetic field is   $B_\mu=(1/2) \epsilon_{\mu \nu \alpha \rho} u^\nu F^{\alpha \rho}.$ 
 By substituting   $ f_{\sschi}$  with (\ref{f_fd}) in   (\ref{generalform}) we get
\begin{eqnarray}
 {{\cal J}^{ {\ssfd} \mu}_\sschi} &&= \bm[ p^\mu \delta(p^2) + \hbar  \chi  Q   p^\nu (u_\nu B^\mu - B_\nu u^\mu)  \delta^\prime (p^2)  \nonumber\\
 &&   +\hbar  \chi p \cdot u  p^\nu (u_\nu \omega^\mu - \omega_\nu u^\mu)  \delta^\prime (p^2)  +  \hbar  \chi (\omega \cdot p) p^\mu  \delta^\prime (p^2)  \nonumber \\
 &&   + \hbar  \chi \omega^\mu \delta(p^2)  + \hbar \chi Q  \epsilon^{\mu \nu \alpha \beta} p_\nu E_\alpha u_\beta   \delta^\prime (p^2)  \bm] 
 f^{\scriptscriptstyle{FD}}_{\sschi}. 
 \label{Jfd}
\end{eqnarray}
The vorticity tensor is expressed in terms of the fluid vorticity $\omega^\mu$ as $\Omega^{\mu \nu }= \epsilon^{\mu\nu\alpha \rho} u_{\alpha} \omega_\rho .$ One can explicitly show that  $(\tilde{\nabla}^\mu \equiv \tilde{\nabla}^\mu_{\ssbfu})$ 
$$\tilde{\nabla}^\mu   {\cal J}^{\ssfd}_{\sschi \mu} = 0,$$
is satisfied. We made use of (\ref{emuT}) and 
$\partial^\mu \tilde{F}_{\mu \nu}=0$ where  $\tilde{F}_{\mu \nu}=(1/2)\epsilon_{\mu \nu \alpha \rho} F^{\alpha \rho},$ which can be expressed as  $\tilde{F}_{\mu \nu}= B_\mu u_\nu - B_\nu u_\mu + \epsilon_{\mu \nu \alpha \beta} E_\alpha u_\beta .$ 

Although the electric field and magnetic field dependent terms of (\ref{Jfd}) are the same with the ones obtained in \cite{glpww}, the vorticity dependent terms are different as it would be expected. However, both should yield the same currents and energy momentum tensor.
We will demonstrate their equivalence by  calculating the currents and energy-momentum tensor resulting from the solution (\ref{Jfd}).

The vector and axial-vector currents are defined  by
\begin{equation}
\label{cdef}
j^\mu = \int d^4p \ {\cal V}^\mu ,\ \ \  j^\mu_5 = \int d^4p \ {\cal A}^\mu .
\end{equation}
By substituting ${\cal V}^\mu$ and ${\cal A}^\mu$ with 
\begin{equation}
 {\cal V}^\mu = \sum_{\chi}  {\cal J}^{\ssfd\mu}_\sschi , \ \ \ \ 
 {\cal A}^\mu = \sum_{\chi}  \chi {\cal J}^{\ssfd\mu}_\sschi ,
 \label{Vfd,Afd}
\end{equation}
 we obtain 
 \begin{eqnarray}
 j^\mu & = & n u^\mu + \xi_{{\ssB}} B^\mu + \xi \omega^\mu ,
 \label{j_mu} \\
 j^\mu_5 &=& n_{\ssbes} u^\mu + \xi_{\ssBb} B^\mu + \xi_{\ssbes} \omega^\mu.
 \label{j_mub}
 \end{eqnarray}
 The total and axial number densities are acquired as
 \begin{eqnarray}
 n= \frac{\mu}{3 \pi^2 \hbar^3} (\mu^2 + 3 \mu_5^2+\pi^2 T^2),
 \nonumber \\ 
 n_{\ssbes}= \frac{\mu_\ssbes}{3 \pi^2\hbar^3} (3 \mu^2 + \mu_5^2+\pi^2 T^2) .\nonumber 
 \end{eqnarray}
 The chiral magnetic and chiral separation effects are given, respectively, by  the second terms of  (\ref{j_mu}) and (\ref{j_mub}), where 
 the coefficients are calculated as
 \begin{equation}
 \xi_{{\ssB}} =\frac{Q\mu_{\ssbes} }{2  \pi^2 \hbar^2} , \ \ 
 \xi_{\ssBb} =  \frac{Q\mu}{2 \pi^2 \hbar^2}  . \nonumber
 \end{equation}
 Similarly, vorticity  yields  the chiral vortical  and local polarization effects  generated by the last terms in (\ref{j_mu}) and (\ref{j_mub}), with
\begin{equation}
\xi = \frac{\mu  \mu_{\ssbes}}{\pi^2\hbar^2} , \ \  \xi_{\ssbes} = \frac{T^2}{6 \hbar^2} + \frac{\mu^2+{\mu_{\ssbes}}^2}{2  \pi^2 \hbar^2} . \nonumber
\end{equation}

The energy-momentum tensor $T^{\mu\nu}$ is defined through the vector field ${\cal V}^\mu$ and four-momentum $p^\mu$ as
\begin{equation}
T^{\mu\nu} = \frac{1}{2} \int d^4p \ (p^\mu  \ {\cal V}^\nu + p^\nu \  {\cal V}^\mu). \nonumber
\end{equation}
By substituting ${\cal V}^\mu$ with (\ref{Vfd,Afd}) and performing the momentum integrals one acquires
\begin{eqnarray}
T^{\mu\nu} 
&=& \varepsilon  u^\mu u^\nu + P (u^\mu u^\nu - g^{\mu\nu} )+ n_{\ssbes} (u^\mu \omega^\nu + u^\nu \omega^\mu )  \nonumber\\
&&+\frac{Q \xi }{2} (u^\mu B^\nu + u^\nu B^\mu ). \nonumber
\end{eqnarray}
Pressure is $P=\varepsilon/3$ and the energy density is computed as $$\varepsilon = \frac{1}{2 \pi^2} \left( \frac{7 \pi^2 T^4 }{30}+ \pi^2 T^2 (\mu^2+\mu_{\ssbes}^2)+3 \mu^2 \mu_5^2 - \frac{\mu^4+\mu_5^4}{2} \right) .$$  

By employing (\ref{emuT})  one can show that the continuity equations are consistent with the chiral anomaly
$
\partial_\mu j^\mu = 0 , \ 
\partial_\mu j^\mu_5 = - (Q^2/2\pi^2) E \cdot B .
$
Moreover,  energy-momentum conservation can be established  as
$
\partial_\mu T^{\mu\nu} = Q F^{\nu\alpha } j_\alpha.
$

All the results obtained by employing the solution (\ref{Jfd}) coincide with the ones presented in \cite{glpww}. Therefore we can conclude that the modified chiral QKE is consistent with the results established in  relativistic hydrodynamics.

\section{Chiral transport equations}    We can now  proceed to formulate the covariant semiclassical CTE from  (\ref{J0}) and (\ref{generalform}),  by working in the comoving frame of fluid by setting $n_\mu=u_\mu.$ Chiral vector fields are defined by the set of equations  (\ref{1st,0})-(\ref{third,0}).  However, (\ref{J0}) and  (\ref{generalform}) satisfy only two of them. The  remaining equation (\ref{2nd,0}),  yields
\begin{eqnarray}
\tilde{\nabla}_\mu  {\cal J}^\mu_\sschi &=& \delta\left( p^2 + \hbar \chi Q \frac{u_\mu \tilde{F}^{\ssmn}   p_\nu }{u \cdot p}\right)   
 \{ p \cdot \tilde{\nabla}  \nonumber\\
&& + \frac{\hbar \chi Q }{u \cdot p} S^{\mu \nu}E_\mu  \tilde{\nabla}_\nu    - \frac{\hbar \chi}{u \cdot p}  p_\mu \tilde{\Omega}^{\ssmn} \tilde{\nabla}_\nu  \nonumber\\
&&+  \frac{\hbar \chi}{u \cdot p}  (\tilde{\Omega}^{\ssmn} p_\mu u_\nu  ) \Omega^{\sigma \rho} p_\rho \partial^{(p)}_\sigma \} f_\chi =0. \nonumber
\end{eqnarray}
We defined  $\tilde{\Omega}_{\mu \nu }=\frac{1}{2}\epsilon_{\mu \nu \alpha \rho} \Omega^{\alpha \rho}$ and $S^{\mu \nu} \equiv S^{\mu \nu}_\ssbfu. $  Observe that in (\ref{generalform})  there 
is a freedom in  choosing  $ f^{1}_\sschi ,$ which permits us to redefine it as
\begin{equation}
f_\chi^1 \Rightarrow \chi \frac{ S^{\mu \nu} \Omega_{\mu \nu}}{u\cdot p}   f_{\sschi}^0 +{f}_\chi^1. \nonumber
\end{equation}
The first term is similar to the rotation  dependent part of the 
frame independent  equilibrium distribution function given in  \cite{css}.  Although this choice is essential to acquire a well-defined 3D limit, it does not violate the uniqueness of the 3D limit. A simple dimensional analysis show that  another term which is linear in vorticity and $f_{\sschi}^0$ does not exist.
Therefore we define the covariant CTE  as 
\begin{flalign}
\delta\left( p^2 + \hbar \chi Q \frac{u_\mu \tilde{F}^{\ssmn}   p_\nu }{u \cdot p}\right) 
\{ p \cdot \tilde{\nabla} \left( 1+ \hbar \chi  \frac{ S^{\mu \nu} \Omega_{\mu \nu}}{u\cdot p}  \right) \nonumber\\
+ \frac{\hbar \chi Q }{u \cdot p} S^{\mu \nu}E_\mu  \tilde{\nabla}_\nu   
- \frac{\hbar \chi}{u \cdot p}  p_\mu \tilde{\Omega}^{\ssmn} \tilde{\nabla}_\nu  \nonumber\\
+  \frac{\hbar \chi}{u \cdot p}  (\tilde{\Omega}^{\ssmn} p_\mu u_\nu  )  \Omega^{\sigma \rho} p_\rho \partial^{(p)}_\sigma \} f_\sschi=0.
\label{nablaj}
\end{flalign}
To ensure its covariance (frame independence) the distribution function $f_\sschi= f_{\sschi}^0 +\hbar {f}_\chi^1 $ should be chosen to transform appropriately \cite{css,hpy1,hpy2,hsjlz}. The main difference between the CTE obtained in \cite{hpy2, hsjlz} and 
 (\ref{nablaj}) lies in the vorticity dependent terms. Obviously another dissimilarity is the fact that (\ref{nablaj}) is defined in the comoving frame of fluid in contrary to the CTE of \cite{hpy2, hsjlz}  which was claimed to be valid in any frame with velocity $n_\mu .$  The electromagnetic field dependent terms and the mass-shell condition coincide by substituting $\tilde{\nabla}_\mu$ with $\nabla_\mu.$ 
 
A method of obtaining  the  3D TE  which is  correlative to a  four-dimensional TE  is to integrate the latter  over $p_0$ \cite{zh1,zh2}. However, integrating four-dimensional TE naively over $p_0$ does not always yield a well-defined  3D equation. This goal can be  achieved  if  the coefficients of $\partial f_\chi / \partial p^\mu$ integrated over $p_0$ permit writing
\begin{eqnarray}
 \dot{\bm p} \left[ \frac{\partial f_{\sschi}(x,p) }{\partial \bm p}\right]_{p_0= {\cal E}} + 
\dot{\bm p}\cdot \frac{\partial  {\cal E} }{\partial \bm p}\left[\frac{\partial f_{\sschi}(x,  p ) }{\partial p_0}\right]_{p_0= {\cal E}} \nonumber\\
=\dot{\bm p}\  \frac{\partial f_{\sschi}(t,\bm x,    {\cal E}, \bm p) }{\partial \bm p},   \label{pep}
\end{eqnarray}
 where  ${\cal E}$ will be dictated by the delta function of the  relativistic equation.  It is worth noting that physical quantities are defined in terms of integrals over momentum variables both in four- and three-dimensions, so that it is sufficient to relate the transport equations  as 
 $$
 \int d^4p \{  \mathrm{4D\ TE}\} =\int d^3p\{\mathrm{3D \ TE}\}.
 $$
 
 We  will integrate   (\ref{nablaj}) over $p_0$ in
 the frame: $u^\mu=(1,\bm 0)$ and $ \omega^\mu= (0, \bm \omega).$ 
  It is possible to satisfy (\ref{pep})
  by setting
\begin{eqnarray}
\int d^4p \ \delta (p^2)\Big\{ E \cdot \omega \left(\frac{f^{0}_{\sschi}}{p_0}-\frac{1}{2}\frac{\partial f^{0}_{\sschi} }{\partial p_0} \right)
\nonumber\\
+\frac{\omega\cdot  p  E \cdot  p }{p_0^2} \left(\frac{2f^{0}_{\sschi}}{p_0}-\frac{1}{2}\frac{\partial f^{0}_{\sschi} }{\partial p_0} \right) \Big\}=0. \label{conf0} \nonumber
\end{eqnarray}
One  can show that this condition is fulfilled for $f_\sschi^0= f^{\scriptscriptstyle{FD}}_{\sschi},$ which is the natural choice for fermions.
Then we can conclude that \mbox{$\int d^4p \{(\ref{nablaj})\} =\int d^3p\{\ \mathrm{3D \ CTE}\}=0,$} where vanishing of  the latter integrand leads to the following 3D  CTE,   
\begin{equation}
\big( \sqrt{\eta}_{\, s}^{\, \sschi } \frac{\partial }{\partial t} + (\sqrt{\eta} \dot{{\bm x}})^\sschi_s  \cdot \frac{\partial }{\partial \bm{x}} + (\sqrt{\eta}  \dot{\bm p})^\sschi_s \cdot\frac{\partial }{\partial \bm{p}}\big) f_{\sschi, s}^{eq} (t,\bm x,\bm p)=0, \nonumber
\end{equation}
with
\begin{eqnarray}
\sqrt{\eta}_{\, s}^{\, \sschi }   &=& 1 + \hbar s Q \chi  \bm b_s \cdot \bm B,  \label{3e1} \\ 
(\sqrt{\eta}  \dot{\bm x})^\sschi_s &=& \bm v_s^\sschi + \hbar \chi  (\hat{\bm p} \cdot \bm b_s)(s Q \bm B + 2 {\cal E}_s^\sschi \bm \omega )  \nonumber \\
&&+ \hbar s  Q  \chi   \bm E \times \bm b_s -2 \hbar \chi (\bm \omega \cdot \bm b_s) \bm p ,\label{3e2}\\
(\sqrt{\eta} \dot{\bm p})^\sschi_s&=& s Q \bm E + s Q  \bm v_s^\sschi  \times \bm B  + \bm v_s^\sschi \times  {\cal E}_s^\sschi \bm \omega \nonumber\\
&& + \hbar \chi Q^2 \bm b_s (\bm E \cdot \bm B) -2 \hbar  \chi (\bm  \omega \cdot \bm b_s)   \bm p\times {\cal E}_s^\sschi \bm \omega  \nonumber \\
&&- 2 \hbar s Q \chi ( \bm \omega \cdot \bm b_s) \bm p \times  \bm B  . \label{3e3}
\end{eqnarray}
We introduced  the Berry curvature $\bm b_s= s \bm p/2 |\bm p|^3.$ 
Dispersion relation  and the canonical velocity are 
\begin{eqnarray}
{\cal E}_s^\sschi &=&  |\bm p| (1- \hbar s Q \chi  \bm b_s \cdot \bm B), \label{disper}\\
\bm v_s^\sschi  &=& \frac{\partial  {\cal E}_s^\sschi }{\partial \bm p} = \hat{\bm p}(1+2 \hbar s  Q \chi   \bm b_s \cdot \bm B)- \hbar s Q  \chi   b_s \bm B. \nonumber
\end{eqnarray}
The third term of (\ref{3e3}) is the  Coriolis force. It  is one half of the force that  one  naively expects to find  by substituting  mass with ${\cal E}_s^\sschi $ in the classical Coriolis force of a massive particle.  

The 3D formalism of \cite{dky} is different from (\ref{3e1})-(\ref{disper}). One of the main differences is in the dispersion relation: (\ref{disper}) does not possess a vorticity dependent term in contrary to the one given in \cite{sy}, \cite{dky}. 
Although the vorticity dependent dispersion relation can be generated by shifting $\bm p$ in  (\ref{disper}),
it is  an open question if 	the current   formalism and the one given in \cite{dky} are  related by a  phase space coordinate transformation  as it is discussed in \cite{hs}. 
	
By making use of  (\ref{3e1}) and  (\ref{3e2}) we can define the chiral
particle (antiparticle) number and current densities as
\begin{eqnarray}
n^\chi_s& = &  \int [dp] (\sqrt{\eta})_s^\sschi f^{eq,s}_{\sschi},\label{nil} \\
\bm j^\chi_s & = & \int [dp](\sqrt{\eta} \dot{\bm x})^\sschi_s f^{eq,s}_{\sschi}+
\bm \nabla \times  \int  [dp]  {\cal E}_s^\sschi  \bm b_s^\sschi f^{eq,s}_{\sschi} ,\label{jil}
\end{eqnarray}
where  $[dp]=d^3p/(2\pi\hbar)^3. $ One can easily observe that they satisfy the continuity equation
\begin{equation} 
\label{ceqD0}
\frac{\partial n_s^\sschi}{\partial t} + \bm {\nabla} \cdot \bm j_s^\sschi= \frac{\chi Q^2}{(2\pi\hbar)^2}  \bm{{ E}}\cdot \bm{B} \ f^{eq,s}_{\sschi}|_{\bm p =0}   . \nonumber
\end{equation}
 The equilibrium partition function of a rotating fluid  \cite{css} in the frame which we integrated (\ref{nablaj})   is
\begin{equation}f^{eq,s}_{\sschi} = \frac{1}{ e^{({\cal E}_s^\sschi - s\mu_\sschi -\hbar s \sschi   \hat{\bm p} \cdot \bm \omega /2)/ T } +1  } .\nonumber \end{equation}
By employing it
one can show that the current (\ref{jil}) yields 
the chiral magnetic   and the  chiral separation effects
$
	\bm j_\ssV^\cme =\xi_{{\ssB}} \bm B, \ 
	\bm j_\ssA^{\ssC \ssS \ssE} =\xi_{\ssBb} \bm B,
$
the chiral vortical  and the local (spin) polarization effects
$
	\bm j_\ssV^\cve =\xi  \bm \omega ,\
	\bm j_\ssA^{\ssL \ssP \ssE} =   \xi_{\ssbes} \bm \omega.
$
They are consistent with  (\ref{j_mu}) and (\ref{j_mub}).

\section{Conclusions} 
  
QKE for fermions is modified to take into consideration the non-inertial properties of the rotating  coordinate frame.
We have constructed its solution for chiral fermions  by choosing  the distribution function as Fermi-Dirac and   calculated  the vector, axial-vector currents and energy momentum tensor. We showed that they are consistent with the hydrodynamical approach which is a necessary condition for being an acceptable theory.  We defined a new  4D CTE. We established its 3D limit which is shown to yield correct anomalous effects as well as continuity equation for number and current densities.  It is a new CKT which is not explicitly coordinate dependent and accommodates the Coriolis force. Various physical systems can be studied within these covariant and 3D transport equations. 
For example they are extremely useful in investigating the nonlinear transport properties of   chiral plasma (see \cite{ofdek} and the references therein).  Incorporating collisions into the covariant CKT  can be addressed by the methods developed in \cite{css,hpy1,hpy2}.
 An important aspect of the modified QKE which  is left  to future studies is to explore its field theoretical origins.

\begin{acknowledgments}
	This work is supported by the Scientific and Technological Research Council of Turkey (T\"{U}B\.{I}TAK) Grant No. 117F328.
	
\end{acknowledgments}
\newcommand{\PRL}{Phys. Rev. Lett. }
\newcommand{\PRB}{Phys. Rev. B }
\newcommand{\PRD}{Phys. Rev. D }

\end{document}